\documentclass[a4paper]{spie}

\usepackage{graphicx}
\usepackage{textcomp}

\title{LEXT: a lobster eye optic for Gamow}

\author[a]{Charlotte Feldman}\author[a]{Paul O'Brien}\author[b]{Nicholas White}\author[c]{Wayne Baumgartner}\author[c]{Nicholas Thomas}\author[a]{Alexander Lodge}\author[d]{Marshall Bautz}\author[d]{Erik Hinrichsen}
\affil[a]{University of Leicester, University Road, Leicester, LE1 7RH, UK}
\affil[b]{The George Washington University, 1918 F Street, NW Washington, DC 20052, USA}
\affil[c]{NASA Marshall Space Flight Center, Martin Road SW, Huntsville, AL 35808, USA}
\affil[d]{MIT Kavli Institute for Astrophysics and Space Research, Massachusetts Institute of Technology, 77 Massachusetts Avenue, Cambridge, MA 02139, USA}

\authorinfo{Further author information: Charly Feldman: \\E-mail: chf7@le.ac.uk, Telephone: +44 (0)116 252 5084}
 
\begin{document} 
\maketitle

\begin{abstract}
The Lobster Eye X-ray Telescope (LEXT) is one of the payloads on-board the Gamow Explorer, which will be proposed to the 2021 NASA Explorer MIDEX opportunity. If approved, it will be launched in 2028, and is optimised to identify high-z Gamma Ray Bursts (GRBs) and enable rapid follow-up. The LEXT is a two module, CCD focal plane, large field of view telescope utilising Micro Pore Optics (MPOs) over a bandpass of 0.2 - 5 keV. The geometry of the MPOs comprises a square packed array of microscopic pores with a square cross-section, arranged over a spherical surface with a radius of curvature of 600 mm, twice the focal length of the optic, 300 mm. Working in the photon energy range 0.2 - 5 keV, the optimum L/d ratio (length of pore L and pore width d) is 60, and is constant across the whole optic aperture. This paper details the baseline design for the LEXT optic in order to fulfil the science goals of the Gamow mission. Extensive ray-trace analysis has been undertaken and we present the development of the optic design along with the optimisation of the field of view, effective area and focal length using this analysis. Investigations as to the ideal MPO characteristics, e.g. coatings, pore size, etc., and details of avenues for further study are also given.
\end{abstract}


\section{INTRODUCTION}
\label{intro}

\subsection{Gamow Explorer}
\label{gam}
The Gamow Explorer is being proposed to the NASA Medium Class Explorer (MIDEX) 2021 opportunity for launch to an L2 orbit in 2028. The Gamow Explorer will be optimized to specifically search for high redshift long Gamma Ray Bursts (LGRBs), with a z $>$ 6 detection rate at least ten times that of Swift\cite{gam}. In addition, by using the photo-z technique, it will autonomously identify LGRBs to enable rapid follow up by large ground based telescopes and JWST. In order to detect as many high z GRBs and other transient events as possible, a large field of view X-ray instrument is required, making a lobster eye telescope an ideal candidate for this mission. Gamow Consists of two instruments, the Photo-z Infra-Red Telescope (PIRT) and the Lobster Eye X-ray Telescope (LEXT), to detect LGRBs and to precisely determine their position and the redshift. The mission is named in honour of George Gamow for his work on and contribution to the Big Bang theory of the expanding Universe. Further details of the mission and science are given in White et al. \cite{gamspie}.

\subsection{Lobster eye telescopes}
\label{lob}
The lobster eye geometry for X-ray imaging was first introduced by Angel (1979)\cite{ang} and the use of tesselated slumped Micro Channel Plates (MCPs) in a lobster eye X-ray telescope has been pursued by several authors\cite{theory, wilks, fraspie, kaa}. This geometry can provide a very large field of view and in addition, due to the low density of the glass used in the optics, the completed telescope has very low mass. This technology and geometry was used on the MIXS-C instrument on BepiColombo\cite{mixs}, is currently being exploited for multiple X-ray telescope missions such as SVOM \cite{gotz}, SMILE\cite{smile} and Einstein Probe\cite{ep}, and proposed missions such as TAP\cite{tap}, LEXI\cite{lexi} and others.

\begin{figure}
	\centering
		\includegraphics[width=0.35\textwidth]{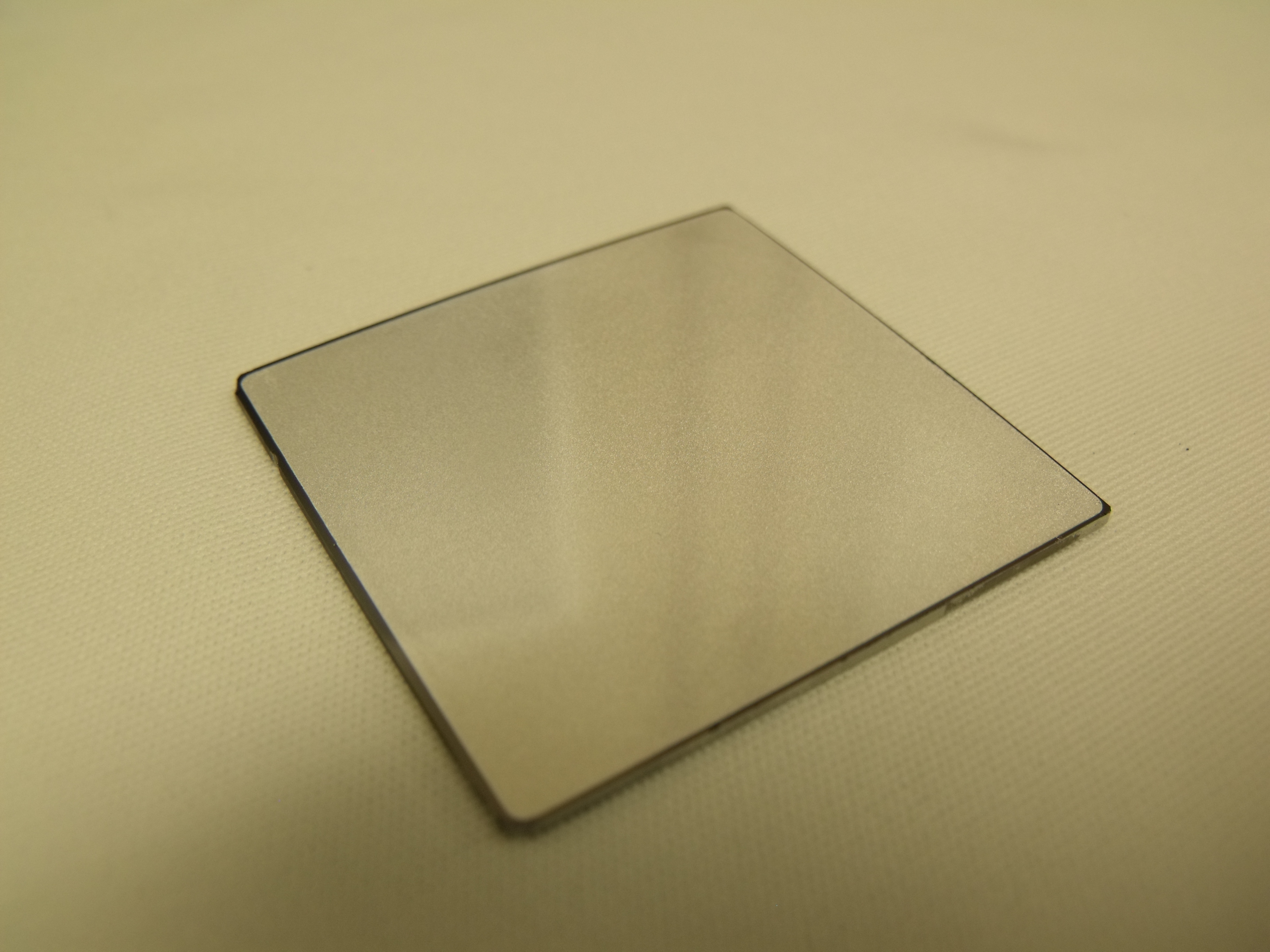}
		\includegraphics[width=0.35\textwidth]{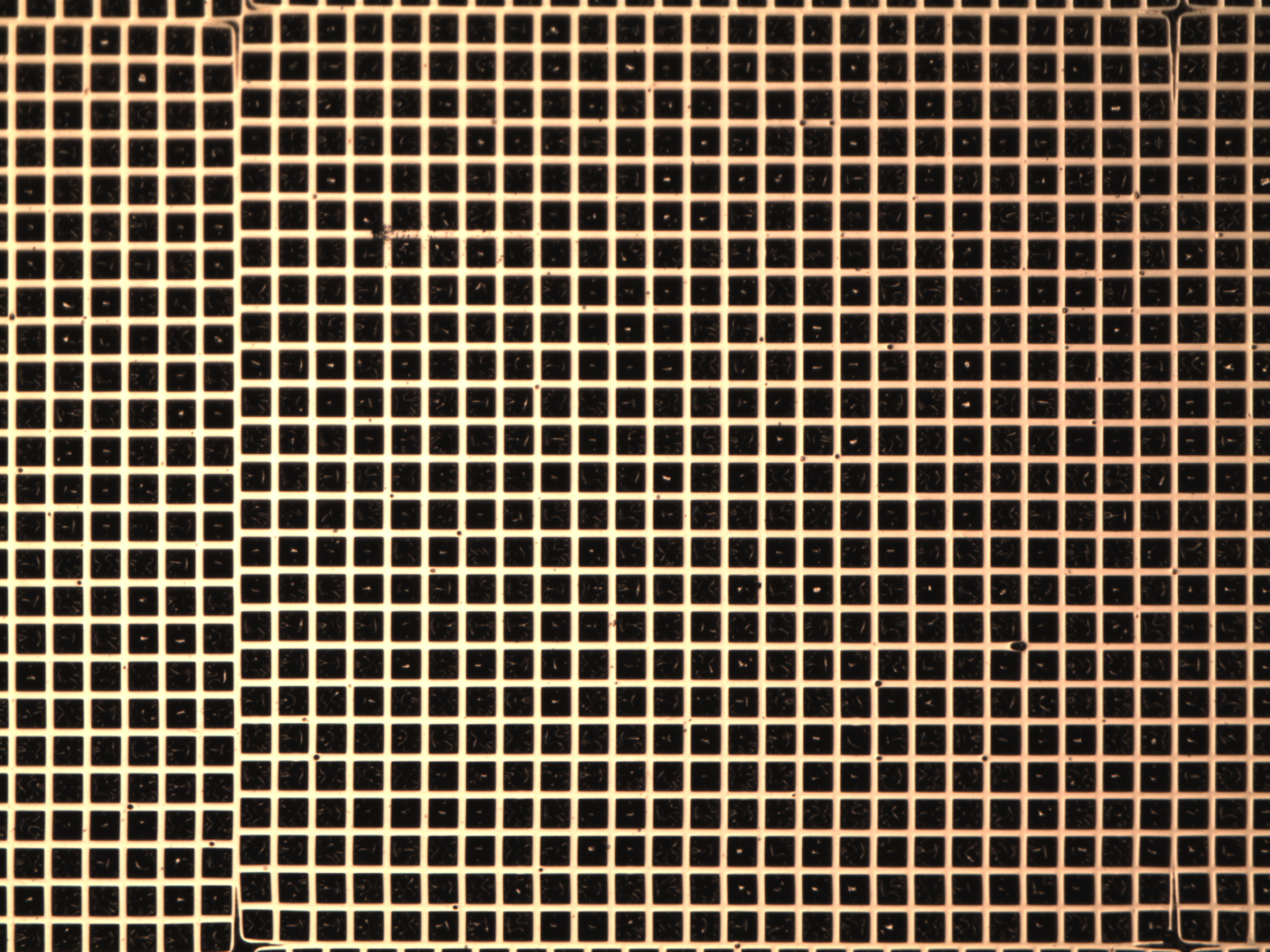}
			\caption{Left: a single, 40 mm by 40 mm square, iridium coated, aluminium filmed, 1.2 mm thick, spherically slumped MPO Produced by Photonis France SAS. Right: a microscope image of a series of individual 40 $\mu$m pores on an MPO.}
	\label{fig2}
\end{figure}

As described in Feldman et al. \cite{svomspie}, a lobster eye optic can be created by tessellating a large number of individual Micro Pore Optics (MPOs) over a spherical frame. The geometry of the individual MPO comprises a square packed array of microscopic pores of a square cross section. The current baseline is for the MPOs to have an iridium coating within the pores and a thin aluminium film applied over the front pore apertures. This provides both a filter for optical and UV light, and a good thermal surface. The MPOs are slumped such that all the pores in all the MPOs point towards the centre of curvature. An example of a slumped, 40 mm by 40 mmm square, 1.2 mm thick, iridium coated, aluminium filmed, spherically slumped MPO is shown in Figure \ref{fig2}.

\begin{figure}
	\centering
		\includegraphics[width=0.6\textwidth]{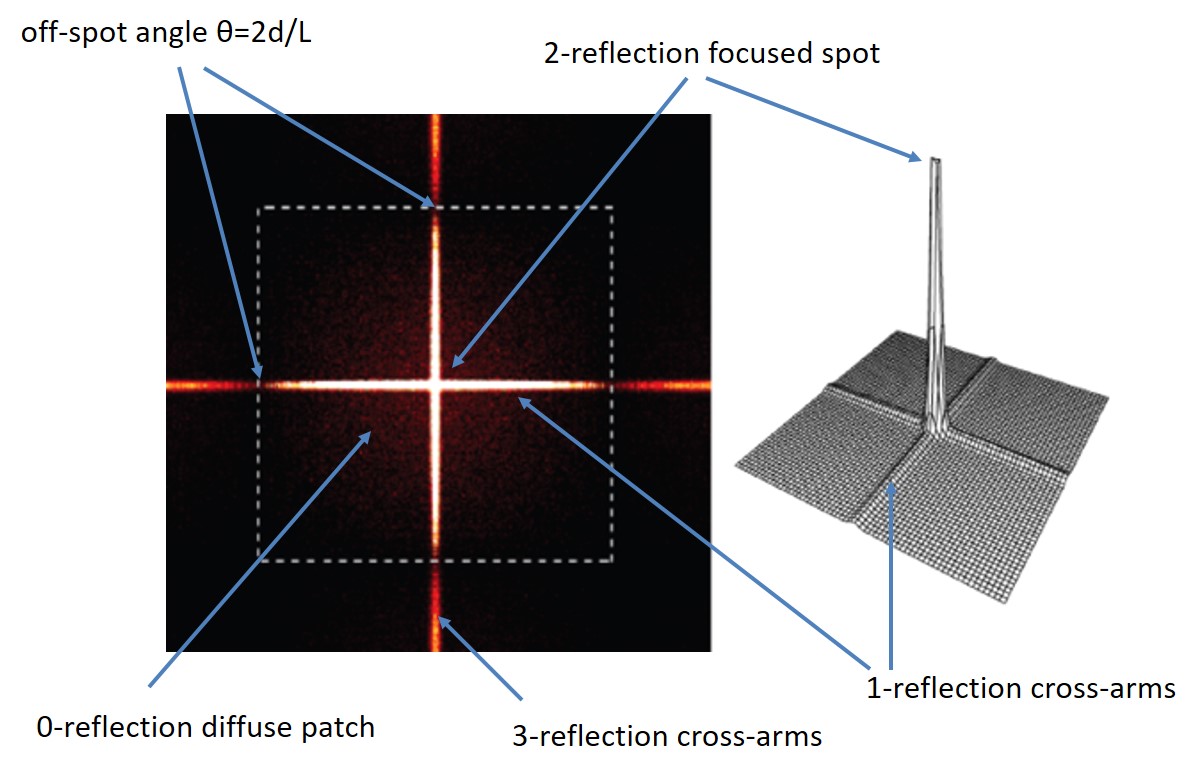}
			\caption{The simulated distinctive PSF created by a perfect, slumped MPO\cite{dickspie}. Rays which have undergone two reflections off orthogonal pore walls produce a high intensity central focussed spot. Rays which undergo a single reflection in the pores produce the horizontal and vertical cross-arms. Three reflection rays and higher contribute to the outer wings and the rays which go straight through the MPO create a diffuse background patch.}
	\label{geom2}
\end{figure}

As shown in Willingale et al. \cite{dickspie}, the point spread function (PSF) produced by a single MPO comprises of a focused spot, created by rays which undergo 2 grazing incidence reflections off orthogonal walls of a pore, vertical and horizontal cross-arms, caused by rays which undergo 1 reflection, and a diffuse patch created by rays which pass straight-through the MPO. An example of the distinctive PSF is shown in Figure \ref{geom2}.

An assembled lobster eye optic consisting of multiple MPOs should produce the same image as shown in Figure \ref{geom2}, if the MPOs have all been mounted without any additional tilts or rotations and the frame has a consistent form over the whole array\cite{svomspie}. If there are any mounting or assembly errors, the central double reflection spot will be broadened or, in extreme conditions, will consist of many individual smaller spots as the Point Spread Functions (PSFs) of the individual MPOs won't pile up on top of each other. In this case there will also be broadening of the cross arms and multiple arms may also be seen.

The size of the field of view of a lobster eye optic depends only on the angular extent of the spherical optic and detector.

\section{THE LEXT}
\label{lext}
\begin{figure}
	\centering
		\includegraphics[width=0.35\textwidth]{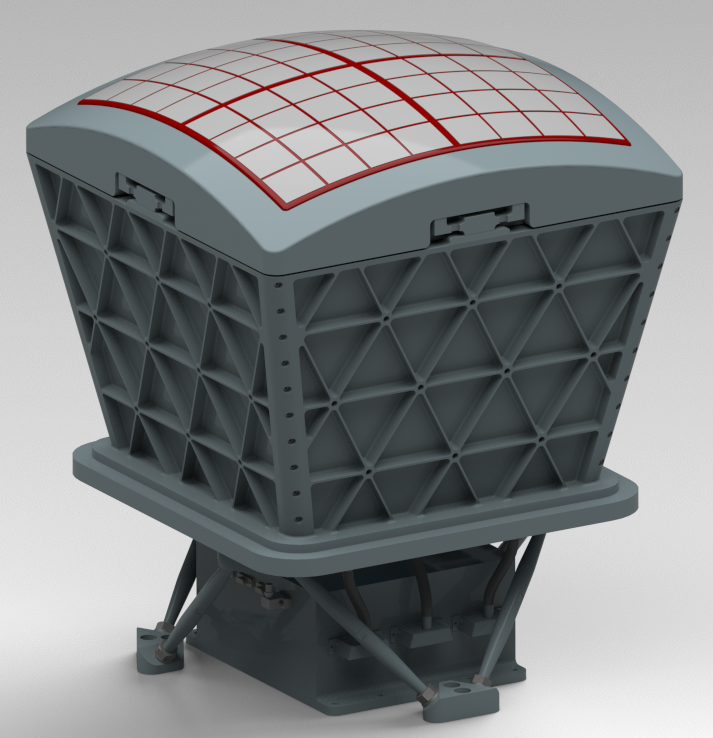}
			\caption{A CAD rendering of a single THESEUS SXI module.}
	\label{the}
\end{figure}

In order to fulfil the science requirements, detect as many high z GRBs as possible, and to provide a large field of view, a multi module lobster eye telescope was proposed. The initial design was based on that of the THESEUS SXI instrument\cite{pobspie} (Figure \ref{the}), but due to the difference in size of the detectors selected for Gamow compared to THESEUS, it was necessary to change the shape of each module.

Gamow will be utilizing CCD detectors originally developed by MIT Lincoln Laboratory for the ISS-TAO mission\cite{tao}, which have 48 mm by 48 mm square active area. Each detector is a 3216 x 3216 array of 15 $\mu$m x 15 $\mu$m pixels, which are binned on-chip to provide an effective pixel size of 45 $\mu$m by 45 $\mu$m.  These devices are four-side abuttable, in principle allowing arbitrary focal plane arrangements, but ease of assembly and filling factor of the field of view are maximized with the elongated focal plane and multiple module approach adopted for Gamow, with the size of the optic calculated to match the size of the focal plane.

\subsection{Design}
\label{opdes}

Two designs are currently being investigated, the first as the baseline design, the second as the threshold design. Both designs are using the same CCD detectors, and both designs are aimed at providing a large field of view and to operate in the energy range of 0.2 - 5 keV. The MPOs are currently assumed to be 2.4 mm thick with an L/d of 60, 40 mm by 40 mm square in size, iridium coated and aluminium filmed. A focal length study was performed and it was determined that a 300 mm focal length (600 mm radius of curvature) provided the best balance between sensitivity and field of view. Values for the surface roughness inside the channels, reflection efficiency and production deformations are all based on the results of previous modelling\cite{dickspie} and experiments completed at the University of Leicester. In both cases the optics size has been determined to match the focal plane size and optimise the field of view.

\begin{figure}
	\centering
		\includegraphics[width=0.40\textwidth]{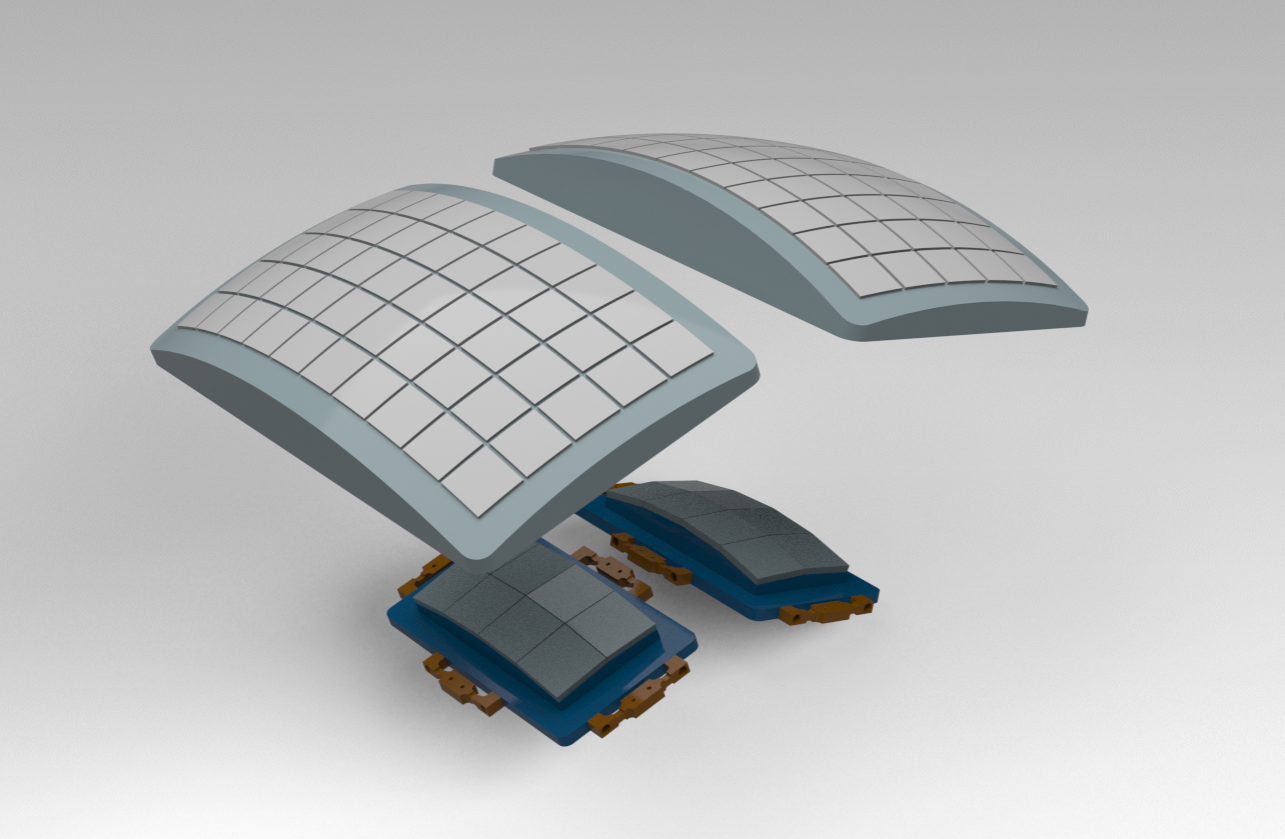}
		\includegraphics[width=0.42\textwidth]{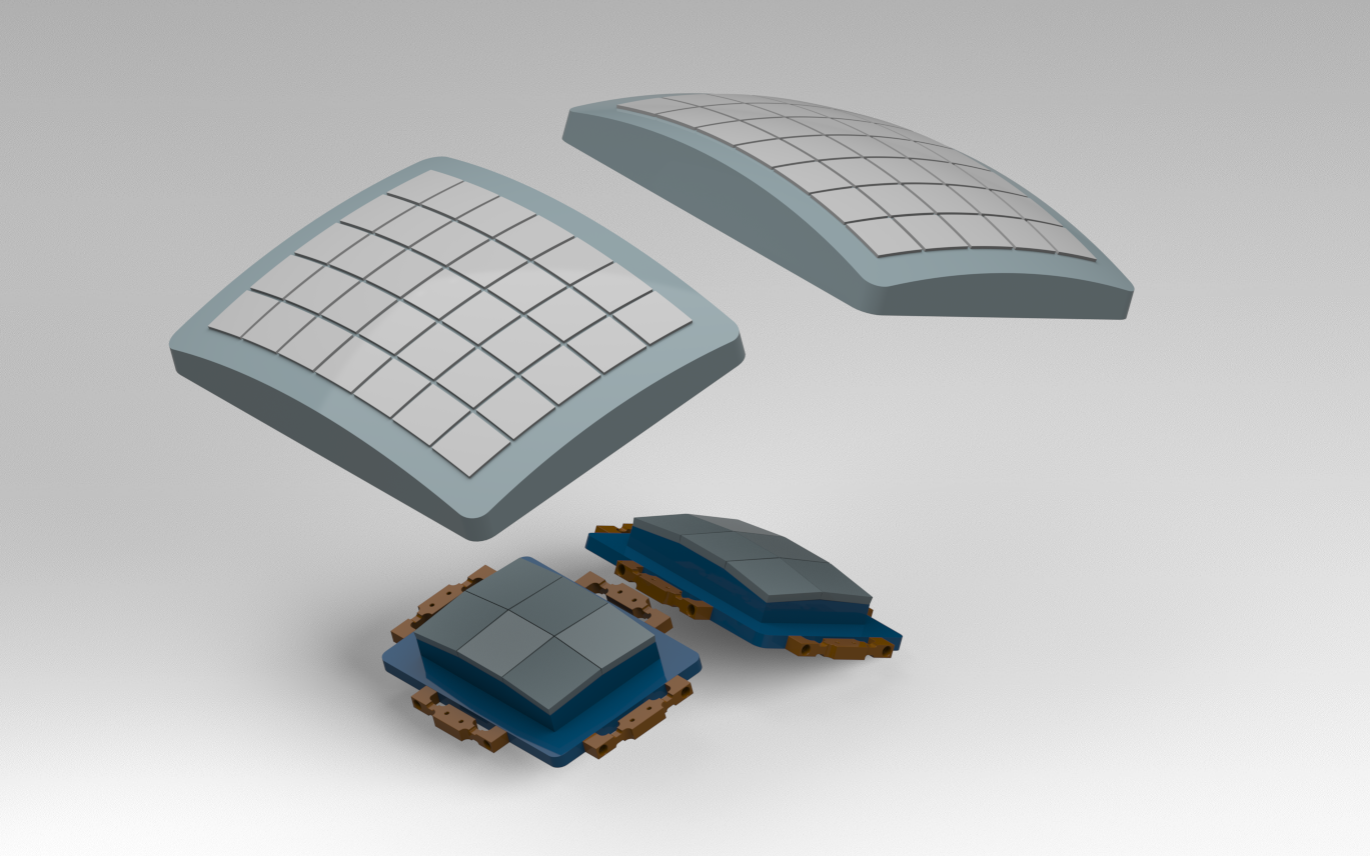}
			\caption{The baseline LEXT design on the left and the threshold deign on the right currently being investigated.}
	\label{design}
\end{figure}

\begin{figure}
	\centering
		\includegraphics[width=0.35\textwidth]{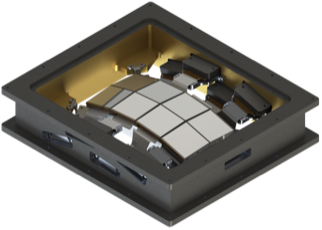}
			\caption{A CAD rendering of the baseline module's focal plane.}
	\label{lfp}
\end{figure}

\begin{flushleft}
\textbf{Baseline design}\\
\end{flushleft}
The baseline design utilises two identical modules with an elongated optic and focal plane. The focal plane consists of a 4 x 2 array of CCD detectors and the optic comprises an array of 10 x 5 MPOs. Each module provides a field of view of $\sim$680 square degrees and the whole instrument has a total field of view of $\sim$1360 square degrees. The two module design is shown on the left of Figure \ref{design} and a CAD rendering of the baseline module's focal plane is shown in Figure \ref{lfp}.

\begin{flushleft}
\textbf{Threshold design}\\
\end{flushleft}
The threshold design is the minimum field of view that would still fulfil the science goals of the mission, namely the number of high z GRBs detectable. Again, two identical modules with an elongated optic and focal plane are used. However, both the number of detectors and MPOs have been reduced. The focal plane comprises an array of 3 x 2 CCDs and each optic is an array of 7 x 5 MPOs. Each module has a field of view of $\sim$500 square degrees with the whole instrument providing just over 1000 square degrees. The threshold design is shown on the right of Figure \ref{design}.

\subsection{Effective area}
\label{ea}
The software used to model the optic designs was created in the sequential ray tracing code Q, designed by Prof. Richard Willingale. Using all the deformations for the individual MPOs, as described in Willingale et al. \cite{dickspie} within Q, it is possible to accurately predict the response of the individual MPOs at different energies and angles. It is also possible to model an array of MPOPs, either by putting each individual MPOs measured properties into the model or by setting an array of identical MPOs with the size and layout desired, as used here. The deformations are randomised using a Gaussian with the peak being set within the code to make the model more realistic as the MPOs won't be exactly identical. By editing the peak values of each deformation type and including assumed frame tilts and assembly errors, it is possible to specify the resultant Full Width Half Maximum (FWHM) of the module and thus calculate the efficiency and sensitivity of different designs and module performances.

\begin{figure}
	\centering
		\includegraphics[width=0.65\textwidth]{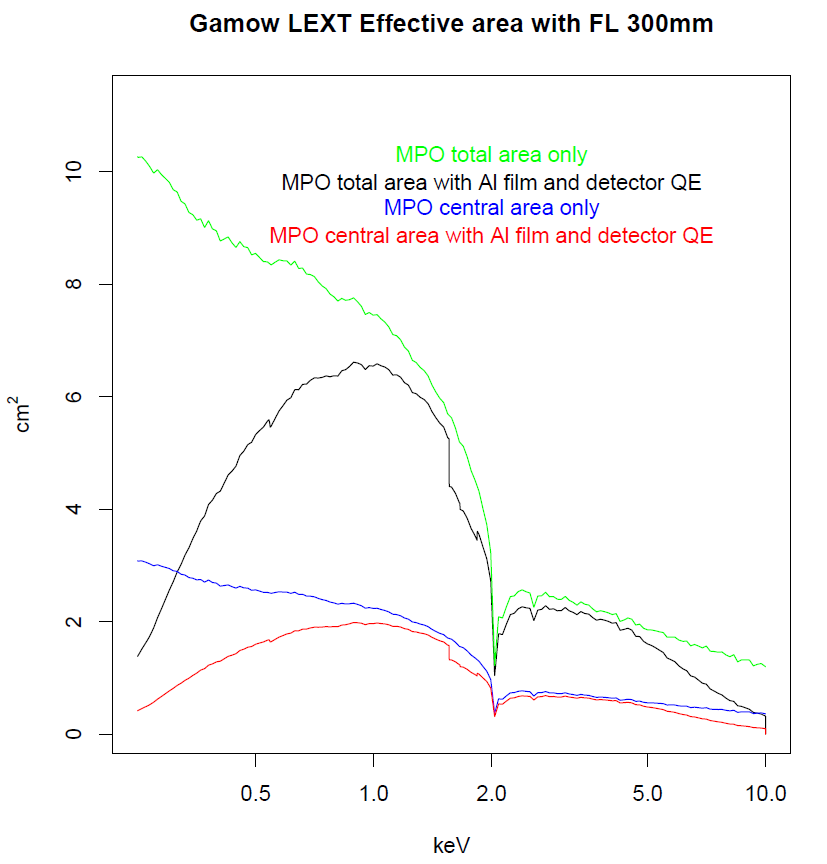}
			\caption{The predicted effective area of the LEXT baseline design with a module FWHM of 7 arcmin.}
	\label{effar}
\end{figure}

Models of the baseline design were ray-traced as described above, and the module FWHM was set to be 7 arcmin. The effective area of the optic, both the central area and the full image, was then measured at energies within an extended bandpass of 0.2 - 10 keV. The optic effective area was then convolved with the detector QE and an aluminium filter (170 nm thick, split between the optic and detector). No other filter, which may be included later on the detector, has been incorporated in the model at this point. The results can be seen in Figure \ref{effar}.

As can be seen in the results, there is some detected area above 5 keV but it does rapidly tail off above this. The central area of the cross beam peaks at 1 keV with $\sim$2 cm$^{2}$.

\section{Further work and studies}
\label{fw}
The current baseline is to have iridium coated pores in the MPOs, however, other materials are being considered in order to be able to improve the optic response across the bandpass.

The pore size for the MPOs is currently assumed to be 40 $\mu$m, however the possibility of having smaller pores of 20 $\mu$m is being considered. This would reduce the thickness of the MPOs from 2.4 mm to 1.2 mm, thus making production slightly faster, but possibly making coating of the pores harder. This trade off is still being investigated.

The sensitivity of the optic is being calculated in order to predict how many high-z GRBs would be detected, which incorporates the current baseline design, detector QE, filters and background count and particle rates. The best filter materials and thicknesses to be used on both the optic and detector at L2 are still being defined. In addition, the background count rate is still being determined based on the measurements made by other X-ray missions at L2. As these values become better known, the predicted sensitivity of the optic can be established.

\section{CONCLUSION}
\label{sumcon}
The Gamow explorer is a NASA MIDEX proposal, which if selected will be launched in to an L2 orbit in 2028. It will consist of two instruments; a GRB detector (LEXT) and an infra-red telescope (PIRT). A lobster eye optic is a perfect candidate for the GRB detector on board Gamow, supplying a large field of view, high sensitivity at detecting GRBs and other transients and provides good localisation. Two designs are currently being studied; a baseline design and a threshold design, both providing over 1000 square degrees field of view over two modules. Each of the two designs incorporates an elongated focal plane of CCD detectors and matched optic size to optimise the field of view. The baseline optic has been modelled and is predicted to provide a peak effective area at 1 keV of $\sim$2 cm$^{2}$. The coating for the pores and the pore size are still being investigated in order to optimise MPO production and the performance of the optic across the bandpass. Preliminary sensitivity calculations have been completed in order to be able to predict the number of high z GRBs that Gamow will be able to detect.

\section{ACKNOWLEDGEMENTS}
\label{ack}
The authors which to thank the entirety of the extensive Gamow team.

This research used the SPECTRE High Performance Computing Facility at the University of Leicester.

\bibliography{bobs}{}
\bibliographystyle{spiebib}
\end{document}